\documentclass[prd, preprint, nofootinbib]{revtex4}
\usepackage{graphicx}
\usepackage{amsfonts}
\usepackage{latexsym}
\usepackage{amssymb}
\usepackage{epsfig}

\begin{document}

\title{
 Isospin violating dark matter being asymmetric
}

\author{Nobuchika Okada}
 \email{okadan@ua.edu}
 \affiliation{
Department of Physics and Astronomy, 
University of Alabama, Tuscaloosa, Alabama 35487, USA
}

\author{Osamu Seto}
 \email{seto@physics.umn.edu}
 \affiliation{
 Department of Life Science and Technology,
 Hokkai-Gakuen University,
 Sapporo 062-8605, Japan
}

%

\begin{abstract}

The isospin violating dark matter (IVDM) scenario 
 offers an interesting possibility to reconcile conflicting 
 results among direct dark matter search experiments 
 for a mass range around 10 GeV. 
We consider two simple renormalizable IVDM models 
 with a complex scalar dark matter and a Dirac fermion dark matter, 
 respectively, whose stability is ensured 
 by the conservation of ``dark matter number.'' 
Although both models successfully work as the IVDM scenario 
 with destructive interference between effective 
 couplings to proton and neutron, 
 the dark matter annihilation cross section 
 is found to exceed the cosmological/astrophysical upper bounds. 
Then, we propose a simple scenario to reconcile the IVDM scenario 
 with the cosmological/astrophysical bounds, 
 namely, the IVDM being asymmetric. 
Assuming a suitable amount of dark matter asymmetry has been
 generated in the early Universe, 
 the annihilation cross section beyond the cosmological/astrophysical upper bound  
 nicely works to dramatically reduce the antidark matter 
 relic density and as a result,
 the constraints from dark matter indirect searches are avoided. 
We also discuss collider experimental constraints 
 on the models and an implication to Higgs boson physics.

\end{abstract}


\preprint{HGU-CAP-021} 

\vspace*{3cm}
\maketitle


\section{Introduction}

Light weakly interacting massive particles (WIMPs) 
 with a mass around 10 GeV have been currently 
 a subject of interest, motivated by some recent results 
 in direct dark matter (DM) detection experiments. 
DAMA/LIBRA has claimed detections of the annual modulation 
 signal by WIMPs~\cite{DAMALIBRA}. 
CoGeNT has found an irreducible excess~\cite{CoGeNT}
 and annual modulation~\cite{CoGeNTan}. 
CRESST has observed many events that expected backgrounds 
 are not enough to account for~\cite{CRESSTII,Brown:2011dp}.
However, these observations are challenged to the null results 
 obtained by other experimental collaborations, 
 CDMS~\cite{CDMS}, XENON10~\cite{XENON10},
 XENON100~\cite{XENON100:2011,XENON100:2012},
 and SIMPLE~\cite{SIMPLE}.

Light WIMPs have been investigated 
 for a dark matter interpretation of those data. 
For instance,
 very light neutralino in the minimal supersymmetric standard model 
 (MSSM)~\cite{Hooper:2002nq,Bottino:2002ry} 
 and the next-to-MSSM (NMSSM)~\cite{Cerdeno:2004xw,Gunion:2005rw}
 or very light right-handed sneutrino~\cite{Cerdeno:2008ep,Cerdeno:2011qv} 
 in the NMSSM.
On the other hand, the Fermi-LAT Collaboration has derived 
 constraints on an $s$-wave annihilation cross section 
 of a WIMP based on the analysis of gamma ray flux~\cite{dSph}. 
Annihilation modes of a light WIMP is now severely constrained.

The isospin violating dark matter (IVDM)~\cite{Feng:2011vu} 
 has been proposed as a way to reconcile the tension 
 between inconsistent results among the direct DM detection 
 experiment, since different nuclei for target material 
 have been used in the detector of each experiments.
The possible consistency between DAMA, CoGeNT~\cite{CoGeNT}, and
 XENON~\cite{XENON10,XENON100:2011} was pointed out~\cite{Feng:2011vu}, 
 while the discrepancy between CoGeNT and CDMS cannot be resolved 
 by IVDM because both of them use germanium as the target.  
However, recently it was reported~\cite{Agnese:2013rvf} that 
 CDMS-II Si have observed three events and its possible signal region overlaps
 with the possible CoGeNT signal region analyzed by Kelso {\it et al.}~\cite{Kelso:2011gd}.
The fitting data with IVDM have been examined 
 by several groups~\cite{Kelso:2011gd,Frandsen:2011ts,Schwetz:2011xm,Farina:2011pw,McCabe:2011sr,Chen:2011vda,Gao:2011bq,Frandsen:2011gi,Frandsen:2013cna},
 and constraints from indirect~\cite{Kumar:2011dr} 
 and direct~\cite{Jin:2012jn} DM detection experiments 
 also have been derived.

In this paper, we consider two simple IVDM models 
 with a complex scalar DM and a Dirac fermion DM, respectively. 
In most of the previous works, the IVDM models have been proposed 
 by introducing a new $U(1)$ gauge symmetry with $Z'$
boson~\cite{Cline:2011zr,Gao:2011ka,Frandsen:2011cg} or an extension 
 of the Higgs sector~\cite{Gao:2011ka,Kawase:2011az,DelNobile:2011yb}.
In contrast to those models, 
 to realize the different cross sections
 with respect to up quarks and down quarks, 
 we introduce fourth generation quarks in the scalar DM model
 and scalar quarks in the fermion DM model, respectively.
Our models are similar to a model briefly mentioned 
 in Ref.~\cite{Feng:2011vu}.

The paper is organized as follows. 
In the next section,
 we describe our models of scalar and fermion DMs. 
In Sec.~III, we identify the allowed region 
 of the mass and couplings of the mediator quarks 
 or scalar quarks by imposing the condition 
 of the isospin violating elastic scattering cross 
 section with nuclei. 
In Sec.~IV, we calculate the annihilation cross section 
 of the IVDM to examine the resultant thermal relic density 
 as well as the constraint from Fermi-LAT data 
 for the parameter region found in Sec.~III. 
Constraints from collider experiments are 
 discussed in Sec.~V. 
Section~VI is devoted to conclusions.

\section{Models}
\subsection{
Model of scalar dark matter with fermion mediators (model S)}

First, we consider a simple model 
 with a complex scalar dark matter, 
 whose particle contents are given in Table~\ref{Table:modelS}. 
In addition to the Standard Model (SM) particle contents, 
 we have introduced the SM $SU(2)$ singlet Dirac fermions
 ($U$ and $D$) whose representations are the same 
 as $SU(2)$ singlet up and down quarks, 
 a complex scalar DM ($\phi$), and a real scalar $S$, 
 with a global $U(1)_G$ symmetry. 
The stability of $\phi$ is ensured by 
 the global $U(1)_G$ symmetry assumed to be conserved. 
All the SM particles are neutral under the global symmetry. 

\begin{table}[t]
\caption{Particle contents for the model S}
\centering
\begin{center}
\begin{tabular}{|c|c|c|c|c|} \hline
${\rm Fields}$ & $SU(3)_c$ & $SU(2)_L$ & $U(1)_Y$ & $U(1)_G$ \\ \hline\hline
$U$            & ${\bf 3}$ & ${\bf 1}$ & $+2/3$   & $+1 $    \\ \hline
$D$            & ${\bf 3}$ & ${\bf 1}$ & $-1/3$   & $+1 $    \\ \hline
$\phi$         & ${\bf 1}$ & ${\bf 1}$ & $0$      & $+1 $    \\ \hline
$S$            & ${\bf 1}$ & ${\bf 1}$ & $0$      & $0  $    \\ \hline
\end{tabular}
\end{center}
\label{Table:modelS}
\end{table}

The gauge and global symmetric Lagrangian 
 relevant to our discussion is given by
\begin{eqnarray}
 {\cal L}  \supset 
   - M_U \overline{U} U -M_D \overline{D} D 
  - \left( 
 f_U \overline{U_L} \phi u_R +  f_D \overline{D_L} \phi d_R 
 + {\rm H.c.} \right)
-V(H, \phi, S), 
\end{eqnarray}
 where $H$ is the SM Higgs doublet, 
 $u_R$ ($d_R$) is the SM right-handed up (down) quark singlet, 
 and $V$ is a scalar potential for $H$, $\phi$, and $S$.

We assume a suitable scalar potential for our discussion:
 not only the Higgs doublet but also the scalar $S$ 
 develop vacuum expectation values 
 and we expand these scalar fields as  
\begin{eqnarray}
 H = \left( 
            \begin{array}{c}
            0 \\
            \frac{1}{\sqrt{2}}(v + h) 
            \end{array}
              \right) , 
~~S = v_s + s ,
\end{eqnarray}
with the vacuum expectation values, $v= 246$ GeV and $v_s$.

After the electroweak symmetry breaking, 
 the SM singlet scalar and the Higgs boson have  
 a mass mixing such that 
\begin{eqnarray}
 \left( 
            \begin{array}{c}
            s \\
            h 
            \end{array}
 \right)
   =
\left( 
            \begin{array}{cc}
            \cos\alpha & \sin\alpha   \\
            -\sin\alpha & \cos\alpha     
            \end{array}
 \right) 
   \left( 
            \begin{array}{c}
            h_1 \\
            h_2 
            \end{array}
  \right)  ,
\end{eqnarray}
where $h_1$ and $h_2$ are the mass eigenstates 
 with masses $m_{h_1} \leq m_{h_2}$, respectively. 
The existence of a light scalar particle mixed 
 with the SM Higgs boson is constrained 
 by the LEP experiments~\cite{LEP1, LEP2}. 
We consider a small mixing, for example, 
 $\sin \alpha < 0.1$, 
 so that the mass eigenstate $h_1$ ($h_2$) 
 is almost the SM singlet scalar (the SM Higgs boson). 
For such a small mixing, the lower mass bound on $h_1$ 
 disappears, and in the following analysis we consider 
 $m_{h_1} < 10$ GeV. 
Terms in the scalar potential relevant to our analysis  below 
 are triple scalar couplings parametrized as
\begin{eqnarray} 
 V  &\supset &  v (\lambda_1 h_1 + \lambda_2 h_2 ) 
 \phi^\dagger \phi + \lambda_3 v h_1^2 h_2 ,
\label{triple-s}
\end{eqnarray}
 with dimensionless couplings $\lambda_{1,2,3}$.
Since the SM-like Higgs boson $h_2$ can decay to 
 the lighter scalars, $h_1$ and $\phi$ 
 ($h_1$ subsequently decays to lighter SM particles), 
 the couplings $\lambda_{2,3}$ should be small 
 in order not to significantly alter the Higgs boson 
 branching ratio from the SM prediction. 
To simplify our analysis, we assume $\lambda_2 \gg \lambda_3$ 
 and further parametrize $\lambda_{1,2}$ 
 as $\lambda_1=\lambda \cos \alpha$ and $\lambda_2=\lambda \sin \alpha$ 
 with $\lambda= \sqrt{\lambda_1^2+ \lambda_2^2}$. 
We will discuss a phenomenological constraint on these parameters 
 from the invisible decay branching ratio 
 of the SM Higgs boson in Sec.~\ref{subsec:HiggsInvisible}.

\subsection{
Model of fermion dark matter with scalar mediators (model F)}

Next, we consider a simple model with a Dirac fermion DM, 
 whose particle contents are given in Table~\ref{Table:modelF}. 
In addition to the SM particle contents, 
 we introduce color triplet scalars 
 ($\tilde{Q}_L$, $\tilde{U}_R$, and $\tilde{D}_R$) 
 that are analogous to the scalar quarks in the MSSM, 
 and a Dirac fermion DM ($\psi$). 
Similarly to the model S, a global $U(1)_{\rm G}$ symmetry 
 has been introduced to ensure the stability of 
 the Dirac fermion DM.  
All the SM fields are neutral under the global symmetry. 

\begin{table}[t]
\caption{Particle contents for the model F}
\centering
\begin{center}
\begin{tabular}{|c|c|c|c|c|} \hline
${\rm Fields}$  & $SU(3)_c$ & $SU(2)_L$ & $U(1)_Y$ & $U(1)_G$ \\ \hline\hline
$\tilde{Q}_L$  & ${\bf 3}$ & ${\bf 2}$ & $+1/6$   & $+1 $    \\ \hline
$\tilde{U}_R$  & ${\bf 3}$ & ${\bf 1}$ & $+2/3$   & $+1 $    \\ \hline
$\tilde{D}_R$  & ${\bf 3}$ & ${\bf 1}$ & $-1/3$   & $+1 $    \\ \hline
$\psi$         & ${\bf 1}$ & ${\bf 1}$ & $0$      & $-1 $    \\ \hline
\end{tabular}
\end{center}
\label{Table:modelF}
\end{table}

The relevant part of the Lagrangian is given by
\begin{eqnarray}
 {\cal L} & \supset & 
 - m_\psi \bar{\psi} \psi 
 - M_Q^2 \tilde{Q}_L^{\dagger} \tilde{Q}_L 
 - M_U^2 \tilde{U}_R^{\dagger} \tilde{U}_R
  -M_D^2 \tilde{D}_R^{\dagger} \tilde{D}_R  \nonumber \\
 &&  
 + A_U \tilde{Q}_L^{\dagger} \tilde{H} \tilde{U}_R
 + A_D \tilde{Q}_L^{\dagger} H         \tilde{D}_R 
 + {\rm H.c.} \nonumber \\
 && 
  - f_L \bar{\psi} \tilde{Q}_L^{\dagger}  q_L 
  - f_{R_u} \bar{\psi} \tilde{U}_R^{\dagger}  u_R 
  - f_{R_d} \bar{\psi} \tilde{D}_R^{\dagger}  d_R + {\rm H.c.}, 
\label{ModelF-int}
\end{eqnarray}
 where $\tilde{H} = i \sigma_2 H^*$, 
 $\tilde{Q}_L=(\tilde{U}_L \, \tilde{D}_L)^T$, 
 $q_L=(u_L \, d_L)^T$ is the SM doublet quark of the first generation, 
 and $A_{U,D}$ are parameters with mass-dimension one.

After the electroweak symmetry breaking, 
 the mass eigenstates of $\tilde{U}$ are obtained as  
\begin{eqnarray}
 \left( 
 \begin{array}{c}
 \tilde{U}_L \\
 \tilde{U}_R
 \end{array}
  \right)  
 = \left( 
 \begin{array}{cc}
  \cos\theta_u  & \sin\theta_u \\
 -\sin\theta_u  & \cos\theta_u
 \end{array}
  \right) \left( 
 \begin{array}{c}
 \tilde{U}_1 \\
 \tilde{U}_2
 \end{array}
  \right) , 
\end{eqnarray}
 with a mixing angle $\theta_u$.
Similarly, $\tilde{D}_1$ and $\tilde{D}_2$ are obtained
 with an angle $\theta_d$.
With the mass eigenstates, 
 the Yukawa interactions between the dark matter fermion 
 and the SM quarks in Eq.~(\ref{ModelF-int}) are rewritten as
\begin{eqnarray}
 {\cal L}_{\rm Y} & = & 
  - \bar{\psi} ( f_L \cos\theta_u P_L - f_{R_u} \sin\theta_u P_R ) 
 \tilde{U}^{\dagger}_1 u
  - \bar{\psi} ( f_L \sin\theta_u P_L + f_{R_u} \cos\theta_u P_R ) 
 \tilde{U}^{\dagger}_2 u \nonumber \\
&-& \bar{\psi} ( f_L \cos\theta_d P_L - f_{R_d} \sin\theta_d P_R ) 
 \tilde{D}^{\dagger}_1 d
  - \bar{\psi} ( f_L \sin\theta_d P_L + f_{R_d} \cos\theta_d P_R ) 
\tilde{D}^{\dagger}_2 d
+ {\rm  H.c.} 
\end{eqnarray}

\section{Dark matter elastic scattering with nuclei}

The dark matter scattering cross section with nucleus ($N$)
 made of $Z$ protons ($p$) and $A-Z$ neutrons ($n$) 
 is given by
\begin{equation}
\sigma_{\rm SI}^N = \frac{1}{\pi}
 \left(\frac{m_N }{m_N + m_{\phi}}\right)^2 ( Z f_p + (A-Z) f_n )^2, 
 \label{sigmaSI:scalar}
\end{equation}
 for a scalar dark matter, while for a Dirac fermion dark matter 
\begin{equation}
\sigma_{\rm SI}^N = \frac{1}{\pi}
 \left(\frac{m_N m_{\psi}}{m_N + m_{\psi}}\right)^2 ( Z f_p + (A-Z)
f_n )^2.   
 \label{sigmaSI:fermion}
\end{equation}
The effective coupling with a proton $f_p$ and a neutron $f_n$ is expressed,
 by use of the hadronic matrix element, as
\begin{equation}
 \frac{f_i}{m_i} = \sum_{q=u,d,s}f_{Tq}^{(i)}\frac{\alpha_q}{m_q} 
  + \frac{2}{27}f_{TG}^{(i)}\sum_{c,b,t}\frac{\alpha_q}{m_q},
\end{equation}
 where $\alpha_q$ is an effective coupling of 
 the DM particle with a $q$-flavor quark 
 defined in the operators
\begin{equation}
{\cal L}_{\rm int}  =
\left\{
\begin{array}{c}
 \alpha_q \bar{q}q |\phi|^2 \qquad {\rm for } \quad \phi   \\
 \alpha_q \bar{q}q \bar{\psi}\psi \qquad {\rm for } \quad \psi 
 \end{array} 
 \right. ,
\end{equation}
 with its mass $m_q$, 
  $f_{Tq}^{(i)}$, and $f_{TG}^{(i)}$ where $ i = p,n $ are constants.
In our analysis, we use the following values: 
 $f_{Tu}^{(p)}=0.0290$, 
 $f_{Td}^{(p)}=0.0352$, 
 $f_{Tu}^{(n)}=0.0195$, 
 $f_{Td}^{(n)}=0.0525$, 
 $f_{Ts}^{(i)}=0$, 
 and $f_{TG}^{(i)}=1-\sum_{q=u,d,s} f_{Tq}^{(i)}$. 
Those $f_{Tu}^{(i)}$ and $f_{Td}^{(i)}$ are quoted from Ref.~\cite{EOS},
 while we set $f_{Ts}^{(i)}=0$ because 
 recent studies of the lattice simulation~\cite{Takeda:2010cw}
 as well as chiral perturbation theory~\cite{Alarcon:2011zs} imply
 negligible strange quark content.
It has been pointed out~\cite{Feng:2011vu} that
 the results of XENON100, CoGeNT and CRESST can be compatible,
 if the following relations are satisfied: 
\begin{eqnarray}
 \frac{f_n}{f_p} \simeq  - 0.7, 
\qquad
 \sigma^p_{\rm SI} \simeq 2 \times 10^{-2}~{\rm pb}.  
\label{IVcondition}
\end{eqnarray}
Note that $f_n \neq f_p$, and therefore the dark matter particle 
 has isospin violating interactions with quarks.

\subsubsection{Model S}

For the model S,  
 there are two contributions to the effective coupling $\alpha_q$. 
One is from the exchange of the scalars $h_1$ and $h_2$, 
 for which we find
\begin{eqnarray}
 \alpha_q &=& - m_q  
           \left( \frac{\lambda_1 \sin\alpha}{m_{h_1}^2} - \frac{\lambda_2 \cos\alpha}{m_{h_2}^2}
           \right) ,
\label{alphaq:scalar1}
\end{eqnarray}
 where we have assumed $m_{h_1}^2 \ll m_{h_2}^2$, 
 with $m_{h_2}$ being the SM(-like) Higgs boson mass. 
Note that $\alpha_q/m_q$ is independent of $q$, 
 so that this contribution conserves the isospin. 
The other contribution is from the exchange 
 of the Dirac fermions, $U$ and $D$: 
\begin{equation}
 \alpha_q = 
 \frac{f_U^2}{2} \frac{m_\phi}{M_U^2-m_\phi^2} \delta^u_q +
 \frac{f_D^2}{2} \frac{m_\phi}{M_D^2-m_\phi^2} \delta^d_q 
 \simeq 
 \frac{f_U^2}{2} \frac{m_\phi}{M_U^2} \delta^u_q +
 \frac{f_D^2}{2} \frac{m_\phi}{M_D^2} \delta^d_q,  
\label{alphaq:scalar2}
\end{equation}
where we have assumed $m_\phi^2 \ll M_{U,D}^2$. 
Clearly this contribution violates the isospin symmetry. 
For simplicity, let us assume $f_D \ll f_U$,  
 and the total contribution is given by 
\begin{eqnarray}
 \alpha_q 
 &\simeq&  - m_q \frac{\lambda_1 \sin\alpha}{m_{h_1}^2} 
 + \frac{f_U^2}{2} \frac{m_\phi}{M_U^2} \delta^u_q \nonumber \\
 &=& - m_q \frac{\lambda \cos\alpha\sin\alpha}{m_{h_1}^2} 
 + \frac{f_U^2}{2} \frac{m_\phi}{M_U^2} \delta^u_q .
\label{alphaq:scalar}
\end{eqnarray} 
Note that the existence of the two terms 
 is crucial to realizing the opposite signs between $f_p$ and $f_n$,
 because the heavy quark $U$ (and also $D$) always positively 
 contributes to $\alpha_q$.

\begin{figure}[t]
\begin{center}
\epsfig{file=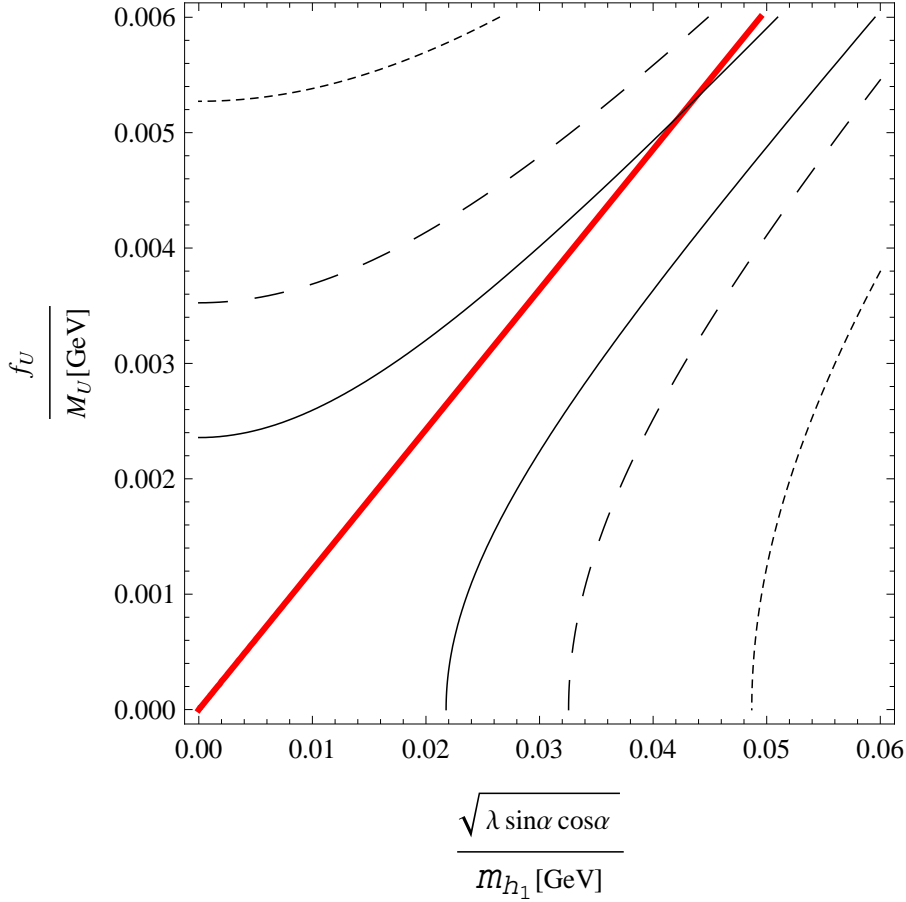, width=12cm,height=12cm,angle=0}
\end{center}
\caption{
 The contours of scattering cross section with a proton 
 for various values, $\sigma^p_{\rm SI}=0.02$ pb (solid), 
 $0.1$ pb (dashed) and $0.5$ pb (dotted), 
 together with the (red) straight line along which 
 the condition $f_n/f_p \simeq -0.7$ is satisfied. 
Here we have fixed the dark matter mass as $m_\phi=8$ GeV. 
 }
\label{SDD}
\end{figure}

Figure~\ref{SDD} shows the contours for various values of 
 $\sigma^p_{\rm SI}$, along with the (red) straight line 
 corresponding to the condition $ f_n/f_p = - 0.7$. 
The two conditions in Eq.~(\ref{IVcondition})
 are satisfied for 
\begin{eqnarray}
 \frac{\sqrt{\lambda \cos\alpha \sin \alpha}}{m_{h_1}} 
 = 4.30 \times 10^{-2} \; {\rm GeV}^{-1} ,~~~~
 \frac{|f_U|}{M_U}    
 =  5.22 \times 10^{-3} \; {\rm GeV}^{-1}. 
 \label{Scalar:param} 
\end{eqnarray}
Here we have fixed the dark matter mass as $m_\phi=8$ GeV.

\subsubsection{Model F}
For the model F, the effective coupling $\alpha_q$ is given by
\begin{eqnarray}
 \alpha_q &=&
- \frac{1}{2} \left[ 
   \sin2\theta_u f_L f_{R_u} 
 \left( \frac{1}{M_{\tilde{U}_1}^2}- \frac{1}{M_{\tilde{U}_2}^2}
 \right) \delta^u_q  
 + \sin2\theta_d f_L f_{R_d} 
 \left( \frac{1}{M_{\tilde{D}_1}^2}- \frac{1}{M_{\tilde{D}_2}^2}
 \right) \delta^d_q
   \right] \nonumber \\ 
&\simeq&
- \frac{1}{2} \left[ 
  \frac{\sin2\theta_u f_L f_{R_u}}{M_{\tilde{U}_1}^2} \delta^u_q  
 +
  \frac{\sin2\theta_d f_L f_{R_d}}{M_{\tilde{D}_1}^2} \delta^d_q
\right].
\label{alphaq:fermion}
\end{eqnarray}
Here, for simplicity, we have taken a limit, 
 $M_{\tilde{U}_1}^2 \ll M_{\tilde{U}_2}^2$ and
 $M_{\tilde{D}_1}^2 \ll M_{\tilde{D}_2}^2$.
This effective coupling violates the isospin symmetry 
 and $f_n/f_p < 0$ can be realized 
 when the relative signs between 
 $\sin2\theta_u f_{R_u}$ and $\sin2\theta_d f_{R_d}$ 
 are opposite. 
We further simplify the system by setting 
\begin{eqnarray}
&& f_L \cos\theta_u = f_{R_u} \sin\theta_u \equiv f_{\tilde U} > 0, 
\nonumber\\
&&
f_L \cos\theta_d = -f_{R_d} \sin\theta_d \equiv f_{\tilde D} > 0 ,
\label{simplify}
\end{eqnarray}
so that 
\begin{eqnarray}
 \alpha_q \simeq 
  - \left(\frac{f_{\tilde U}}{M_{\tilde{U}_1}} \right)^2 \delta^u_q  
  +
  \left( \frac{f_{\tilde D}}{M_{\tilde{D}_1}} \right)^2 \delta^d_q  .
\label{alphaq:fermion2}
\end{eqnarray}

\begin{figure}[h,t]
\begin{center}
\epsfig{file=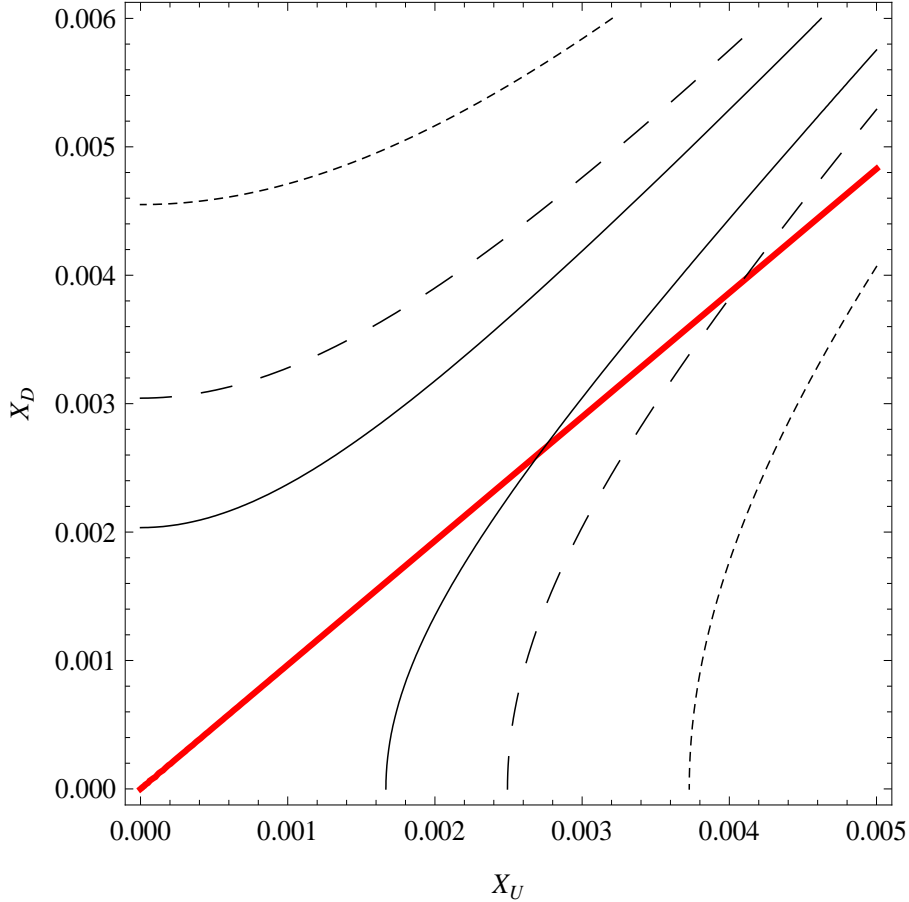, width=12cm,height=12cm,angle=0}
\end{center}
\caption{
The same as Fig.~\ref{SDD} but for the Dirac fermion dark matter. 
Here $X_U \equiv \frac{f_{\tilde U}}{M_{\tilde{U}_1}}$,
 and $X_D \equiv \frac{f_{\tilde D}}{M_{\tilde{D}_1}}$. 
 }
\label{FDD}
\end{figure}

Figure~\ref{FDD} shows the contours for various values of 
 $\sigma^p_{\rm SI}$, along with the (red) straight line 
 corresponding to $ f_n/f_p = - 0.7$. 
The two conditions in Eq.~(\ref{IVcondition})
 are satisfied for
\begin{eqnarray}
\frac{f_{\tilde U}}{M_{\tilde{U}_1}}
=2.73 \times 10^{-3} \; {\rm GeV}^{-1} ,~~~~
\frac{f_{\tilde D}}{M_{\tilde{D}_1}}
=2.63 \times 10^{-3} \; {\rm GeV}^{-1} .
\label{Fermion:param}
\end{eqnarray}

%

\section{Dark matter annihilation cross section}
In this section, we estimate the annihilation cross section 
 of the scalar/fermion dark matter particles 
 for the parameters identified in the previous section 
 to satisfy the conditions for the IVDM. 
%
%
We will see that the $s$-wave annihilation cross section 
 of the dark matter is too large to reproduce 
 the observed relic abundance. 
In order to achieve the correct relic abundance, 
 one may consider a nonthermal dark matter scenario. 
However, this scenario cannot be viable, 
 because the $s$-wave annihilation cross section already exceeds  
 the upper bound obtained by the Fermi-LAT observations~\cite{dSph}. 
In the last part of this section, we will propose 
 a simple scenario to realize the IVDM 
 being consistent with the Fermi-LAT observations.

\subsection{Model S}
The dominant dark matter annihilation process 
 is found to be $\phi^\dagger \phi \to b \bar{b}$ 
 mediated by the scalars, $h_1$ and $h_2$, in the $s$ channel. 
Assuming $m_{h_1}^2 \lesssim m_\phi^2 \ll m_{h_2}^2$, 
 the $s$-wave annihilation cross section is evaluated as 
\begin{eqnarray}
 \langle \sigma v \rangle \simeq 
 \frac{3}{16 \pi} 
\left( 
 \frac{\lambda \sin 2 \alpha \; m_b}{4 m_\phi^2-m_{h_1}^2}
\right)^2, 
\label{XsecS}
\end{eqnarray}
  where $m_b=4.2$ GeV is the bottom quark mass. 
Using the values in Eq.~(\ref{Scalar:param}), 
 we find, for example, $\langle \sigma v \rangle \simeq 6.12$ pb 
 for $m_\phi=8$ GeV and $m_{h_1}=2.9$ GeV. 
This cross section is roughly one order of magnitude 
 larger than the typical dark matter annihilation 
 cross section $\langle \sigma v \rangle \simeq 1$ pb 
 to achieve the observed relic density. 
Thus, in this case, the resultant dark matter 
 abundance becomes too small.   
In order to realize the observed relic density, 
 we may assume a nonthermal production of dark matter particles 
 in the early Universe. 
However, this cannot be a phenomenologically viable scenario, 
 because the dark matter annihilation cross section 
 to the bottom  quarks is constrained 
 by the Fermi-LAT data as 
 $\langle \sigma v \rangle \lesssim 0.5$ pb~\cite{dSph}. 
In this case, the dark matter is overabundant and 
 the relic density should be diluted by some mechanism 
 in the history  of the Universe. 
Since such a scenario is quite ambiguous, 
 we do not consider it in this paper.

\subsection{Model F}

The $s$-wave annihilation modes are given 
 by $t$-channel $\tilde{U}/\tilde{D}$ exchange 
 with $u \bar{u}/d \bar{d}$ final states. 
In a limit 
 $m_{\psi}^2 \ll M_{\tilde{U}_{1,2}, \tilde{D}_{1,2}}^2$, 
 the cross sections is found to be 
\begin{eqnarray}
 \langle \sigma v \rangle \simeq  a_u+ a_d ,    
\label{sigmav:fermion}
\end{eqnarray}
 with
\begin{eqnarray}
 a_u &=& \frac{ N_c m_{\psi}^2}{4\pi }
 \left[ f_L^4
  \left( \frac{\sin^2 \theta_u}{M_{\tilde{U}_2}^2} 
 + \frac{ \cos^2 \theta_u}{M_{\tilde{U}_1}^2} \right)^2 
 + f_{R_u}^4 \left(\frac{\cos^2\theta_u}{M_{\tilde{U}_2}^2} 
 + \frac{\sin^2\theta_u}{M_{\tilde{U}_1}^2} \right)^2   \right. \nonumber \\
&& \left.
 + \sin^2 2\theta_u f_L^2 f_{R_u}^2  
 \left( \frac{1}{M_{\tilde{U}_1}^2}- \frac{1}{M_{\tilde{U}_2}^2}
\right)^2
  \right] 
\simeq
\frac{ 3 N_c m_{\psi}^2}{2\pi } 
 \left( \frac{f_{\tilde U}}{M_{\tilde{U}_1}} \right)^4 \\
 a_d &=& \frac{ N_c m_{\psi}^2}{4\pi }
 \left[ f_L^4
  \left( \frac{\sin^2 \theta_d}{M_{\tilde{D}_2}^2} 
 + \frac{ \cos^2 \theta_d}{M_{\tilde{D}_1}^2} \right)^2 
 + f_{R_d}^4 \left(\frac{\cos^2\theta_d}{M_{\tilde{D}_2}^2} 
 + \frac{\sin^2\theta_d}{M_{\tilde{D}_1}^2} \right)^2   \right. \nonumber \\
&& \left.
 + \sin^2 2\theta_d f_L^2 f_{R_d}^2  
 \left( \frac{1}{M_{\tilde{D}_1}^2}- \frac{1}{M_{\tilde{D}_2}^2}
\right)^2
  \right] 
\simeq
\frac{ 3 N_c m_{\psi}^2}{2\pi } 
 \left( \frac{f_{\tilde D}}{M_{\tilde{D}_1}} \right)^4 ,
\end{eqnarray}
 where we have used Eq.~(\ref{simplify}) 
 and the limit $M_{\tilde{U}_1}^2 \ll M_{\tilde{U}_2}^2$ 
 and $M_{\tilde{D}_1}^2 \ll M_{\tilde{D}_2}^2$. 
Using the values in Eqs.~(\ref{Fermion:param}),
 we find the annihilation cross section as 
\begin{equation}
\langle \sigma v \rangle \simeq 3.68 \; {\rm pb},
\end{equation}
 which is too large to reproduce the correct thermal relic density 
 of the dark matter particle in the present Universe. 
In order to make the relic abundance right, 
 we may consider a nonthermal production of the dark matter particles 
 in the early Universe.  
However, as in the model S, such a scenario is not viable 
 by the Fermi-LAT observations~\cite{dSph}. 
The upper bound on the cosmic antiproton flux 
 obtained by the Fermi-LAT observations 
 is interpreted to a cross section upper bound 
 of DM annihilations to up and down quarks as~\cite{Kumar:2011dr} 
\begin{equation}
\langle \sigma v \rangle \lesssim 0.2 \; {\rm pb}. 
\end{equation}

\subsection{Solution to too large annihilation cross section}
As we have seen, 
 for a given parameter set to realize a large enough isospin violating
 scattering cross section with nuclei, 
 the resultant annihilation cross section is too large to satisfy cosmological
 and astrophysical constraints.
For relic density, one may assume a nonthermal dark matter production.
However, as we have seen, such an idea cannot work 
 because of the severe upper bound on the dark matter annihilation 
 cross section from the Fermi-LAT observations. 
In order to avoid the Fermi-LAT constraints, 
 we propose an extension of our model to 
 the so-called  ``asymmetric dark matter'' scenario~\cite{Barr:1990ca,
 Barr:1991qn,Kaplan:1991ah,Thomas:1995ze,Hooper:2004dc,
 Kitano:2004sv,Kaplan:2009ag}.
This scenario is suitable to our model, 
 because the global $U(1)_G$ symmetry introduced in our model 
 leads to the conservation of the dark matter number. 
Once a suitable DM-anti-DM asymmetry is created 
 in the early Universe, the too large annihilation cross 
 section nicely works to leave only the dark matter 
 in the present Universe with the observed relic abundance. 
Since the relic abundance of antidark matter particles 
 in the present Universe is much smaller than the dark matter one, 
 a cosmic ray flux produced by DM and anti-DM annihilations 
 becomes much smaller and hence the constraint from the Fermi-LAT 
 observations can be avoided.

A relic density of the dark matter particles 
 in the presence of dark matter asymmetry (chemical potential) 
 has been analyzed in detail by solving the Boltzmann equations~\cite{IDC}.
For example, with a suitable initial dark matter asymmetry, 
 the observed relic abundance of the dark matter particle 
 can be obtained by the $s$-wave annihilation cross section 
 $\langle \sigma v \rangle = {\cal O}(1)$ pb, 
 while the relic abundance of antidark matter particle 
 is found to be $2$ orders of magnitude smaller 
 than the dark matter one. 
As annihilation cross sections become larger, 
 the relic abundance of anti-DM particle 
 becomes exponentially smaller. 
This result is almost independent of WIMP dark matter mass. 
We apply the result to our scenario, 
 so that the cosmic ray flux from DM-anti-DM pair 
 annihilations is significantly suppressed 
 and the constraint from the Fermi-LAT observations is avoided.

\section{ Constraints from collider experiments }

\subsection{Constraints on the mediator (s)quarks from LHC}
Our model includes heavy (s)quarks, 
 which can be produced at the Large Hadron Collider (LHC) 
 mainly through the gluon fusion process. 
The heavy (s)quarks, once produced, decay to 
 the SM quarks and the dark matter particles,
 and this process is observed as the hadronic final 
 states with transverse missing energy. 
Searches for such events have been performed at the LHC experiments, 
 and the null result, so far, sets the lower bound 
 on heavy (s)quark masses as $\gtrsim 800$ GeV~\cite{simplifiedSUSY}. 
This bound is obtained for the so-called simplified MSSM, 
 where scalar quarks of the first two generations 
 are produced at the LHC and decay to quarks and 
 the lightest superpartner neutralino. 
Since we only introduced one generation of heavy (s)quarks, 
 the mass bound on the mediator (s)quarks should be a little milder, 
 but let us apply the bound for conservative discussion. 
From Eqs.~(\ref{Scalar:param}) and (\ref{Fermion:param}), 
 we can see that this mass bound is satisfied 
 with the couplings being in a perturbative regime, 
 $f_U^2/(4 \pi), f_{{\tilde U}, {\tilde D}}^2/(4 \pi)\ll 1$.

\subsection{Constraint from Higgs boson invisible decay}
\label{subsec:HiggsInvisible}
In model S, the scalar mass eigenstate $h_2$ 
 is approximately identified as the SM Higgs boson. 
Through the mass mixing with the singlet scalar $s$, 
 the SM Higgs boson decays to a pair of the dark matter particles.~\footnote{
This structure is the same as in the so-called Higgs portal dark 
 matter scenario. See, for example, Ref.~\cite{KMNO} 
 for a detailed analysis and references therein. }
This decay width is given by 
\begin{eqnarray}
 \Gamma(h_2 \to \phi \phi^\dagger) = 
 \frac{\lambda^2 \sin^2\alpha v^2}{16 \pi m_{h_2}}
 \sqrt{1-\frac{4 m_\phi^2}{m_{h_2}^2}}. 
\end{eqnarray}
The current ATLAS~\cite{ATLAS-Higgs} and CMS~\cite{CMS-Higgs} 
 data for the Higgs boson production and its various decay modes 
 are mostly consistent with the SM expectations, 
 and the branching ratio of an invisibly decaying Higgs boson
 is constrained (at 3$\sigma$) as~\cite{Hinv} 
\begin{equation}
 {\rm BR}(h_2 \to {\rm invisible}) 
 = \frac{\Gamma(h_2 \to \phi \phi^\dagger)}
 {\Gamma_{\rm SM} + \Gamma(h_2 \to \phi \phi^\dagger)}
 \leq 0.35,
\label{Gamma_inv}  
\end{equation}
where $\Gamma_{\rm SM}=4.07$ MeV~\cite{HiggsDecay} 
 is the SM prediction of the total decay width 
 of a Higgs boson with a 125 GeV mass.

\begin{figure}[t]
\begin{center}
\epsfig{file=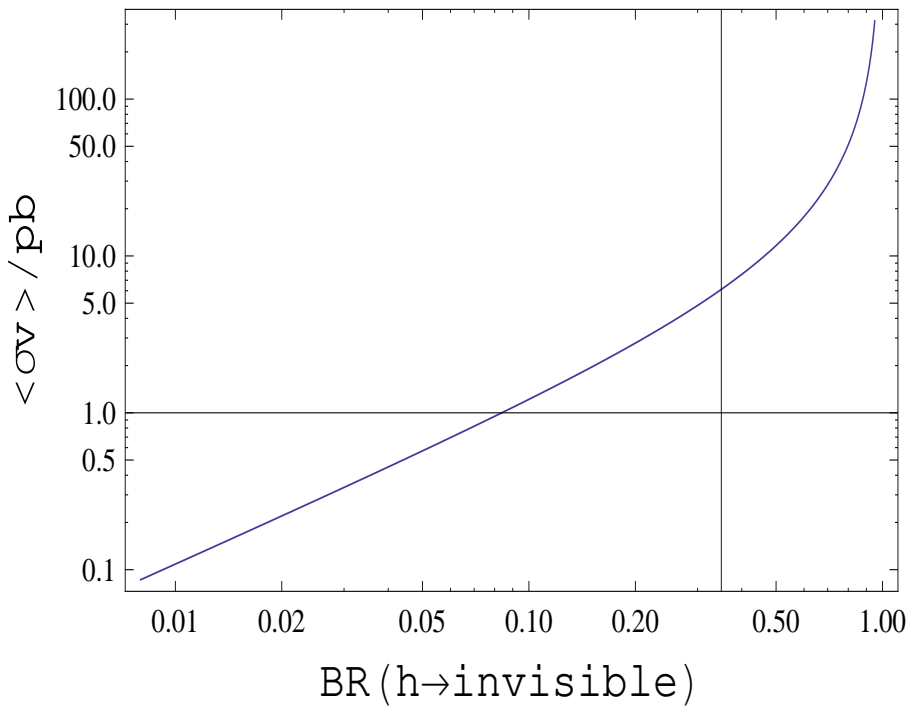, width=12cm, height=10cm,angle=0}
\end{center}
\caption{
The correlation between 
 the Higgs invisible decay rate and 
 the dark matter annihilation cross section 
 through $m_{h_1}$. 
Here we have varied $m_{h_1}$ in the range of 
 $1 \leq m_{h_1}({\rm GeV}) \leq 7.0$.
} 
\label{Fig-Corr}
\end{figure}

Using the result in Eq.~(\ref{Scalar:param}), 
 we can give the annihilation cross section of Eq.~(\ref{XsecS}) 
 and the Higgs invisible decay rate of Eq.~(\ref{Gamma_inv}) 
 as a function of only $m_{h_1}$, 
 with a fixed dark matter mass $m_\phi=8$ GeV. 
The correlation between these two quantities 
 is shown in Fig.~\ref{Fig-Corr} 
 by varying $m_{h_1}$ in the range of $1$ GeV $\leq m_{h_1} \leq 7.0$ GeV. 
Here the vertical line denotes the upper bound, 
 ${\rm BR}(h_2 \to {\rm invisible})=0.35$ 
 at 3$\sigma$~\cite{Hinv}
 while the horizontal line corresponds to 
 a typical value ($\langle \sigma v \rangle =1$ pb) 
 of the WIMP dark matter annihilation cross section 
 for reproducing the observed relic abundance. 
The upper bound ${\rm BR}(h_2 \to {\rm invisible})=0.35$ 
 is obtained by $m_{h_1}\simeq 2.9$ GeV, 
 for which we find the annihilation cross section 
 $\langle \sigma v \rangle \simeq 6.1$ pb.   
Note that the asymmetric IVDM scenario we have proposed 
 in the previous section can be consistent 
 with the constraint on the Higgs invisible decay rate. 
In order for the asymmetric dark matter 
 to be consistent with the observed relic abundance, 
 we have a lower bound on the annihilation cross section  
 as $\langle \sigma v \rangle \gtrsim 1$ pb~\cite{IDC}. 
Applying this bound, we read 
 ${\rm BR}(h_2 \to {\rm invisible}) \gtrsim 8\%$ 
 from Fig.~\ref{Fig-Corr}. 
Precision measurements of Higgs decay width at future 
 collider experiments such as 
 the international linear collider, photon collider and muon collider 
 can reveal the existence of the dark matter.

\section{Conclusions}
The IVDM scenario with destructive interference 
 between the dark matter scatterings with a proton and a neutron 
 offers an interesting possibility to reconcile 
 conflicting results among direct dark matter search experiments 
 for a light WIMP with mass around 10 GeV. 
In this paper, we have considered two simple IVDM models 
 and investigate various phenomenological aspects of the models, 
 such as realization of the IVDM scenario, the constraints 
 on dark matter annihilation cross sections 
 from the dark matter relic abundance as well as 
 an indirect search for dark matter, and collider experimental 
 constraints on the extra particles introduced in our models.

One model introduces a complex scalar as a dark matter particle 
 along with heavy extra quarks and a SM singlet real scalar, 
 through which the dark matter particle couples with 
 the SM up and down quarks. 
Isospin violating effective couplings are realized 
 by the interference between processes mediated by 
 the heavy quarks and the scalar. 
In the other model, we have introduced a Dirac fermion 
 as a dark matter particle along with heavy colored scalars 
 analogous to squarks in the MSSM, 
 through which the dark matter particle couples 
 with the SM quarks. 
The interference between two processes mediated 
 by up-type squarks and down-type squarks 
 realizes the isospin violating effective couplings. 
For both models, we have identified a parameter region 
 suitable for the IVDM scenario. 
With the parameter regions, we have also calculated 
 the relic abundance of the dark matter which is found 
 to be too large to reproduce the observed relic abundance. 
%
%

We have noticed that for both models, the calculated 
 dark matter annihilation cross sections exceed 
 the upper bound obtained by the Fermi-LAT observations too, 
 and therefore the parameter regions for realizing 
 the IVDM scenario are excluded. 
We have proposed a simple scenario to reconcile 
 the IVDM scenario with the Fermi-LAT observations, 
 namely, the IVDM being asymmetric. 
In our models, a global $U(1)_{\rm G}$ symmetry has been 
 introduced whose conservation ensures the stability of 
 a dark matter particle. 
At the same time, this global symmetry leads to 
 the conservation of the dark matter number 
 and this structure is suitable for the asymmetric 
 dark matter scenario. 
As discussed above, we have found that the dark matter annihilation  
 cross section is too large to
 satisfy cosmological and astrophysical constraints simultaneously. 
In fact, when a suitable asymmetry 
 between dark matter-antidark matter is generated 
 in the early Universe, the large cross section nicely 
 works to leave only the dark matter in the present Universe.
Thus, the relic abundance of the antidark matter particle 
 is much less than the dark matter relic abundance; 
 as a result, the flux of cosmic rays created by annihilations
 of the dark matter and antidark matter particles 
 is dramatically suppressed and the constraint 
 by the Fermi-LAT observations is avoided.

Since a variety of models to account for generating 
 the dark matter asymmetry has been proposed 
 (for an incomplete list, 
  see e.g.,~\cite{An:2009vq, Haba:2010bm,Blennow:2010qp, 
  Falkowski:2011xh, Okada:2012rm, Bell:2011tn,
 Cheung:2011if, vonHarling:2012yn, MarchRussell:2011fi, 
 Kamada:2012ht,Unwin:2012rp}),
 we do not propose a specific model for it in this paper. 
However, we should note that some ``dark matter number violating''
 operator,
 in other words the global $U(1)_G$ breaking terms, 
 is necessary to generate the dark matter asymmetry 
 in the Universe and such an operator might induce 
 a dark matter number violating mass term at low energies,
 which must be sufficiently suppressed~\cite{Buckley:2011ye} 
 not to spoil the asymmetric dark matter scenario. 
Concretely speaking, in model F, for instance, 
 we may introduce the following scenario
 by means of a scalar condensate, which is
 analogous to the Affleck-Dine baryogenesis~\cite{AD}. 
Although none of scalar fields carrying $U(1)_G$ charges  
 develop vacuum expectation values at the present Universe, 
 we may add the global $U(1)_G$ as well as the $CP$ violating potential,
 which is given as a function of the gauge invariant product, 
 $\tilde{U}_R \tilde{D}_R \tilde{D}_R$~\footnote{
 To be precise, this product should be like 
 $\tilde{U}_R\tilde{D}_R\tilde{S}_R$, 
 where an extra flavor of an additional down-type 
 scalar quark ($\tilde{D}_R$) has been introduced. },
 in the scalar potential.
During the time that the Universe undergoes a false vacuum 
 with nonvanishing expectation value~\footnote{
 Here, we assume that our scalar potential has an appropriate form
 so that this direction is flat enough to develop an expectation value
 in the early Universe, while the existence of such a flat 
 direction ($D$-flat direction) is automatic 
 for the original Affleck-Dine mechanism 
 in the context of supersymmetric models.} 
 of $\langle \tilde{U}_R\tilde{D}_R\tilde{D}_R \rangle$,
 dark matter asymmetry can be dynamically generated 
 through the evolution of the coherent scalar 
 in the similar way as the Affleck-Dine baryogenesis~\cite{AD}.
Note that although the global $U(1)_G$ symmetry is explicitly broken 
 by terms with $\tilde{U}_R \tilde{D}_R \tilde{D}_R$, the model still possesses 
 a residual $Z_3$ symmetry under which we may assign charges as 
 $\tilde{Q}_L: \omega$, $\tilde{U}_R: \omega$, 
 $\tilde{D}_R: \omega$, $\psi: \omega^2$, 
 where $\omega =e^{i 2 \pi/3}$. 
This $Z_3$ symmetry forbids a Majorana mass term 
 for the dark matter. 
As above,
 in order not to induce the dark matter number violating mass term,
 the $U(1)_G$ breaking should arise via operators that respect a $Z_N$
 subgroup of $U(1)_G$, with $N \geq 3$,
 independently of what mechanism actually generates the asymmetry.
Then, this $Z_N$ symmetry forbids a dark matter number violating mass term 
 for the dark matter. 
%
%
%
%

We have also considered collider experimental constraints on our model. 
Colored fermions and scalars introduced in our models  
 can be produced at the LHC and their decays to 
 the SM quarks and dark matter particles yield the signal 
 events with jets and missing transverse energy. 
We have confirmed that our IVDM scenario 
 is realized consistently with the current LHC bound 
 on the mass of the colored particles. 
In the model S, the SM Higgs boson invisibly decays to 
 a pair of dark matter particles and the upper bound on 
 the invisible decay rate is given by the LHC data. 
We have found a parameter region in which the IVDM 
 scenario is consistent with the LHC bound on 
 the Higgs boson invisible decay rate. 
Interestingly, our successful asymmetric IVDM scenario 
 leads to a lower bound on the invisible decay rate about 8\%, 
 so that precision measurements of the Higgs decay width 
 at future collider experiments can test our scenario.

Observable effects of the asymmetric dark matter scenario 
 in neutron stars have been investigated~\cite{ADM-BH1,ADM-BH2}. 
Since the dark matter particles do not self-annihilate, 
 once captured in neutron stars, dark matter particles 
 are continuously accumulating and neutron stars 
 eventually collapse into black holes. 
Observations of old neutron stars provide constraints 
 on parameters of the asymmetric dark matter scenario. 
In particular, such constraints are more severe 
 for the case with a scalar dark matter 
 because of the absence of Fermi degeneracy pressure. 
However, since the resultant constraints highly depend on 
 the strength of dark matter self-interactions~\cite{ADM-BH2}, 
 we do not consider the constraints from the black  
 formation in our scenario.

%
%
\section*{Acknowledgments}
This work was supported in part by 
 the DOE Grant No. DE-FG02-10ER41714 (N.O.), 
 and by the scientific research grants from Hokkai-Gakuen (O.S).
O.S would like to thank the Department of Physics and Astronomy 
 at the University of Alabama for their warm hospitality 
 where this work was initiated.



\end{document}